\documentclass[aps,prd,twocolumn,titlepage,superscriptaddress,nofootinbib]{revtex4}
\usepackage{graphicx}
\usepackage{color}
\usepackage{latexsym}
\usepackage{cancel}

\usepackage[normalem]{ulem}
\usepackage{hyperref,amssymb}
\usepackage{url}
\usepackage{color,soul}
\usepackage{graphicx}
\usepackage{verbatim}
\usepackage{multirow}
\usepackage{amsmath}
\newcommand{\beq}{\begin{eqnarray}}
	\newcommand{\eeq}{\end{eqnarray}}
\usepackage{mathrsfs}
\usepackage{float,soul}
\usepackage[dvipsnames]{xcolor}
\usepackage{mathtools}
\usepackage{slashed}
\usepackage{physics}	
\usepackage{graphicx}   
\usepackage{epstopdf}
\usepackage{subfigure}  
\usepackage{hyperref}   
\usepackage{bbold}
\usepackage{wasysym}
\usepackage{feynmp}
\usepackage{hyperref}
\hypersetup{colorlinks,citecolor=blue}

\makeatletter
\newcommand{\myBig}{\bBigg@{1.75}}
\makeatother
\begin{document}
	
	\title{Interior structure of black holes with nonlinear terms} 
	\author{Zi-Qiang Zhao}
	\affiliation{Liaoning Key Laboratory of Cosmology and Astrophysics, College of Sciences, Northeastern University, Shenyang 110819, China}
	\author{Zhang-Yu Nie}\email{niezy@kust.edu.cn}
	\affiliation{Center for Gravitation and Astrophysics, Kunming University of Science and Technology, Kunming 650500, China}
	\author{Xing-Kun Zhang}
	\affiliation{Center for the Cross-disciplinary Research of Space Science and Quantum-technologies (CROSS-Q), College of Physics, Nanjing University of Aeronautics and Astronautics, Nanjing, 210016, China}
	\author{Yu-Sen An}
	\affiliation{Center for the Cross-disciplinary Research of Space Science and Quantum-technologies (CROSS-Q), College of Physics, Nanjing University of Aeronautics and Astronautics, Nanjing, 210016, China}
	\author{Jing-Fei Zhang}
	\affiliation{Liaoning Key Laboratory of Cosmology and Astrophysics, College of Sciences, Northeastern University, Shenyang 110819, China}
	\author{Xin Zhang}\email{zhangxin@neu.edu.cn}
	\affiliation{Liaoning Key Laboratory of Cosmology and Astrophysics, College of Sciences, Northeastern University, Shenyang 110819, China}
	\affiliation{MOE Key Laboratory of Data Analytics and Optimization for Smart Industry, Northeastern University, Shenyang 110819, China}
	\affiliation{National Frontiers Science Center for Industrial Intelligence and Systems Optimization, Northeastern University, Shenyang 110819, China}

	\begin{abstract}
		We investigate the oscillation of the Kasner exponent $p_t$ near critical point of the hairy black holes dual to holographic superfluid and reveal a clear inverse periodicity $f(T_c/(T_c-T))$ in a large region below the critical temperature. We first introduce the fourth-power term with a coefficient $\lambda$ to adjust the oscillatory behavior of the Kasner exponent $p_t$ near the critical point. Importantly, we show that the nonlinear coefficient $\lambda$ provides accurate control of this periodicity: a positive $\lambda$ stretches the region, while a negative $\lambda$ compresses it. By contrast, the influence of another coefficient $\tau$ is more concentrated in regions away from the critical point. This work provides a new perspective for understanding the complex dynamical structure inside black holes and extends the actively control from the fourth- and sixth-power term into the black hole interior region.
	\end{abstract}
	\maketitle


	\section{Introduction}
	As one of the most significant astrophysical objects in gravitational physics, black holes remain at the forefront of fundamental research. On the one hand, treating black holes as thermodynamic systems allows us to explore profound connections between quantum gravity and statistical physics~\cite{Bekenstein:1973ur, Hawking:1975vcx}. On the other hand, based on the AdS/CFT correspondence~\cite{Maldacena:1997re}, the holographic properties of black holes provide a powerful and unique theoretical framework for investigating phase transitions such as high-temperature superconductivity~\cite{Hartnoll:2008vx, Hartnoll:2008kx,Herzog:2010vz,Gubser:2008wv,Cai:2013aca,Chen:2010mk,Kim:2013oba,Basu:2010fa,Musso:2013ija,Nie:2013sda,Amado:2013lia,Donos:2013woa,Nie:2015zia,Nie:2014qma,Li:2017wbi,Zhang:2021vwp}, strongly coupled quantum field theories and particularly non-equilibrium processes in condensed matter physics~\cite{Zeng:2022hut,Xia:2023pom,Xia:2021jzh,Zhao:2023ffs,Su:2023vqa,Zeng:2024rwn}.
	
	For a long time, research in black hole physics has predominantly focused on their astrophysical  properties. However, a more fundamental and fascinating question arises: what physical laws govern the internal structure of black holes beyond the event horizon? Recent studies have shown that scalarized black holes exhibit rich dynamical behaviors in their interiors~\cite{Cai:2020wrp,Hartnoll:2020fhc,Wang:2020nkd,An:2021plu,Mansoori:2021wxf,Liu:2021hap,Cai:2021obq,An:2022lvo,Sword:2022oyg,Xu:2023fad,Zhang:2025hkb,Zhang:2025tsa,Xu:2025edz}: the inner horizon disappears and is replaced by a series of dynamical regimes, including the collapse of the Einstein-Rosen bridge, Josephson oscillations, ultimately settling into a Kasner universe. The behavior of the scalar field in the Kasner regime is characterized by the Kasner exponent $p_t$, which exhibits highly oscillatory behavior near the critical point of the scalarization phase transition.
	
	A natural question follows: does this oscillatory structure, which encodes information about the black hole interior, possess deeper regularities? Can its pattern be precisely described and controlled? Meanwhile, introducing nonlinear interaction terms into holographic superfluid models has been shown to significantly alter the system’s phase transition and condensation behavior~\cite{Zhao:2022jvs, Zhao:2024jhs, Zhao:2025tqq}. This motivates us to investigate: could nonlinear terms also exert a crucial influence on the internal structure of black holes, particularly on the oscillatory behavior of the Kasner exponent?
	
	This work aims to systematically study the impact of higher-order nonlinear terms on the interior Kasner geometry of black holes within a holographic s-wave superfluid model. Our numerical calculations find that the nonlinear coefficient $\lambda$ can effectively tune the period of the oscillatory behavior: a positive $\lambda$ stretches the region, while a negative $\lambda$ compresses it. This discovery not only provides a new perspective for understanding the complex periodic structure within black holes but also reveals a novel approach to actively controlling this structure through model parameters. The behavior inside a black hole is indeed rich, and a complete description involves multiple regions extending from the horizon to the singularity. The Kasner exponents provide a relatively simple quantitative measure and exhibit interesting inverse periodic behavior near the critical point. Typically, universal laws near a critical point hold only within an infinitesimally small linear region around it. However, this inverse periodic behavior remains stable over a finite range near the critical point, making it particularly intriguing to explore the influence of nonlinear terms on this phenomenon. Such investigation can deepen our understanding of the critical universality exhibited inside black holes.
	
	This paper is divided into the following sections. In Sect.~\ref{sec2}, we introduce the holographic model and the details of numerical calculations. In Sect.~\ref{sec3}, we present the derivation process and the specific form of the Kasner metric. In Sect.~\ref{sec4}, we provide our numerical results. Finally, we give some conclusions in Sect.~\ref{sec5}.
	
	

	\section{Holographic setup}\label{sec2}
	We first briefly introduce the model we employ. We consider a charged scalar field with additional higher-order nonlinear terms $\lambda(\psi^{\ast}\psi)^{2}$ and $\tau(\psi^{\ast}\psi)^{3}$. When the coefficients of nonlinear term $\lambda$ and $\tau$ are equal to zero, it reduces to the simplest holographic s-wave superconductor model~\cite{Hartnoll:2008vx,Hartnoll:2008kx}. Thus, the total action takes the following form
	\begin{align}
		&S=\,S_{M}+S_{G}~,\quad
		S_G=\,\frac{1}{2\kappa_g ^2}\int d^{4}x\sqrt{-g}\left(R-2\Lambda\right)~,\label{Lagg}\\
		&S_M=\,\frac{1}{2\kappa_g ^2}\int d^{4}x\sqrt{-g}(-\frac{1}{4}F_{\mu\nu}F^{\mu\nu}
		-D_{\mu}\psi^{\ast}D^{\mu}\psi  \nonumber\\
		&~~~~~~~~~~-m^{2}\psi^{\ast}\psi-\lambda(\psi^{\ast}\psi)^{2}-\tau(\psi^{\ast}\psi)^{3})~,\label{Lagm}
	\end{align}
	in which $\Lambda=-3/L^2$ and $\kappa_g^2=8\pi G$, in the rest of this paper, we take $L=1$. Here, $F_{\mu\nu}=\nabla_{\mu}A_{\nu}-\nabla_{\nu}A_{\mu}$ is the Maxwell field strength and $D_{\mu}\psi=\nabla_{\mu}\psi-i A_\mu\psi$ is the standard covariant derivative term of the charged scalar field $\psi$.
	
	The Einstein equation is
	\begin{align}
		R_{\mu\nu}-\frac{1}{2}(R-2\Lambda)g_{\mu\nu}=\frac{1}{2}\mathcal{T}_{\mu\nu}~,
	\end{align}
	where $\mathcal{T}_{\mu\nu}$ is the stress-energy tensor of the matter fields
	\begin{align}
		\mathcal{T}_{\mu\nu}=&(-\frac{1}{4}F_{\alpha\beta}F^{\alpha\beta}
		-D_{\alpha}\psi^{\ast}D^{\alpha}\psi-m^{2}\psi^{\ast}\psi\nonumber\\
		&-\lambda(\psi^{\ast}\psi)^{2}-\tau(\psi^{\ast}\psi)^{3})g_{\mu\nu}+(D_{\mu}\psi^{\ast}D_{\nu}\psi\nonumber\\
		&+D_{\nu}\psi^{\ast}D_{\mu}\psi)+F_{\mu\alpha}F^{\alpha}_{\nu}.
	\end{align}
	The metric and matter ansatz are as follows
	\begin{align}
		&ds^{2}=\frac{1}{z^2}(-f(z)e^{-\chi(z)}dt^{2}+\frac{1}{f(z)}dz^{2}+dx^{2}+dy^{2})~,\\
		&~~~~~~~~~~~~~~~~\psi=\psi(z)~,~A_\mu dx^\mu=\phi(z)dt~.
	\end{align}
	In this case, the temperature of the black hole is defined as
	\begin{align}
		T=\frac{1}{4\pi}f'(z_h)e^{-\chi(z_h)/2}~.
	\end{align}
	Based on the above ansatz, we can obtain the complete equations of motion
	\begin{align}
		&z^2e^{-\chi/2}(e^{\chi/2}\phi')'=\frac{2q^2\psi^2}{f}\phi~,\label{EM1}\\
		&z^2e^{\chi/2}(\frac{e^{-\chi/2}f}{z^2}\psi')'=
		(\frac{m^2}{z^2}\psi-\frac{q^2e^{\chi}\phi^2}{f}\psi\nonumber\\
		&~+\frac{2\lambda}{z^2}\psi^3+\frac{3\tau}{z^2}\psi^5)~,\label{EM2}\\
		&\frac{\chi'}{z}=(\frac{q^2e^{\chi}}{f^2}\phi^2\psi^2+\psi'^2)~,\label{EM3}\\
		&4e^{\chi/2}z^4(\frac{e^{-\chi/2}f}{z^3})'=(2m^2\psi^2+z^4e^{\chi}\phi'^2-12\nonumber\\
		&+2\lambda\psi^4+2\tau\psi^6)~,\label{EM4}
	\end{align}
	in which
	\begin{align}
		f(z)=1-2z^3M(z)~.
	\end{align}
	
	To numerically solve the equations of motion, we need to provide the expansions at the horizon $z\rightarrow z_h$ and at infinity $z\rightarrow 0$. The expansions at the horizon are
	\begin{align}
		\phi(z)=&\phi_{h_1}(z-z_h)+\phi_{h_2}(z-z_h)^2+\cdots~,\\
		\psi(z)=&\psi_{h_0}+\psi_{h_1}(z-z_h)+\cdots~,\\
		\chi(z)=&\chi_{h_0}+\chi_{h_1}(z-z_h)+\cdots~,\\
		M(z)=&\frac{1}{2z^3_h}+M_{h_1}(z-z_h)+\cdots~.
	\end{align}
	Near AdS boundary, the expansion of
	the functions are
	\begin{align}
		\phi(z)&=\mu-z\rho+\cdots~,\\
		\psi(z)&=z\psi^{(1)}+z^2\psi^{(2)}+\cdots~,\\
		\chi(z)&=\chi_{b_0}+z^3\chi_{b_3}+\cdots~,\\
		M(z)&=M_{b_0}+zM_{b_1}+\cdots~.
	\end{align}
	
	In this paper, we work in the canonical ensemble, which means that we fix the charge $\rho$ and chemical potential $\mu$ of the black hole while treating the temperature as a variable of the system.

	\section{Interior solutions of black holes and Kasner geometry}\label{sec3}
	The addition of an extra charged scalar field to the black hole system leads to the disappearance of the inner Cauchy horizon and the formation of a series of dynamical regions~\cite{Cai:2020wrp,Hartnoll:2020fhc}. Beyond the horizon, the metric component $g_{tt}$ undergoes a rapid exponential contraction, a process that destroys the original inner horizon (the collapse of ER bridge). Subsequently, driven by the background gauge potential $A_t$, the charged scalar field exhibits behavior analogous to the Josephson effect in superconductivity. These oscillations leave observable imprints on the metric functions and serve as a key transitional phase connecting the initial collapse to the later Kasner universe; their specific parameters determine the subsequent evolution of spacetime (Josephson oscillation). After the Josephson oscillation, the spacetime geometry settles into a stable Kasner epoch. In this Kasner regime, all spatial metric components evolve as power laws in time (here, either the $z$-coordinate inward from the horizon or the proper time $\xi$), while the scalar field evolves logarithmically. Moreover, the Kasner exponents satisfy the relations that their sum and the sum of their squares both equal one. Ultimately, the final Kasner epoch terminates at the singularity. In some cases, Kasner inversion, transition, alternation behavior may also occur. Detailed analytical analysis has been provided in Ref~\cite{Hartnoll:2020fhc}.
	
	Before diving into numerical analysis, we first need to obtain the explicit form of the Kasner metric. Following Ref.~\cite{Hartnoll:2020fhc}, we can neglect the mass term of the scalar field and the charge term in the Maxwell equation. By ignoring the the higher-order nonlinear terms of scalar field. The Eqs.~(\ref{EM1}) \textbf{--} (\ref{EM4}) then become
	\begin{align}
		&\phi'=E_0e^{-\chi/2},\label{EM1inBH}\\
		&(\frac{e^{-\chi/2}f}{z^2}\psi')'=0,\label{EM2inBH}\\
		&\frac{\chi'}{z}=\psi'^2~,\label{EM3inBH}\\
		&(\frac{e^{-\chi/2}f}{z^3})'=(\frac{1}{4}E_0^2-\frac{12}{z^4})e^{-\chi/2}~,\label{EM4inBH}
	\end{align}
	in which $E_0$ is the constant of the electric field.

	Using Eq.~(\ref{EM2inBH}) {and setting the scalar field as $\psi=\sqrt{2}\int(\Psi/z)dz$, where $\Psi=\alpha$ (analytic solution to Eq.~(\ref{EM2inBH}) for the scalar field; see Ref.~\cite{Hartnoll:2020fhc} for details)}, the solutions of Eqs.~(\ref{EM1inBH}) \textbf{--}  (\ref{EM4inBH}) can be written as
	\begin{align}
		f&=-f_Kz^{3+\alpha^2}+\dots~,\nonumber\\
		\psi&=\alpha\sqrt{2}\log z+\dots~,\nonumber\\
		\chi&=2\alpha^2\log z+\chi_K+\dots~,\label{SolutionSimpleEq}
	\end{align}
	where $f_K$ and $\chi_K$ are constants.
	
	Solving the simplified equations of motion, we can obtain the metric components $g_{tt}$ and $g_{zz}$ in the following form
	\begin{align}
		g_{tt}=f_Kz^{1-\alpha^2},~g_{zz}=-\frac{1}{f_K}z^{-5-\alpha^2}.
	\end{align}
	
	Transforming the z coordinate to the proper time $\xi$, we can obtain the Kasner form of the metric
	\begin{align}
		ds^2&=-d\xi^2+c_t\xi^{2p_t}dt^2+c_x\xi^{2p_x}(dx^2+dy^2)~,\nonumber\\
		\psi&=-p_{\psi}\log\xi~.
	\end{align}
	Here $c_t$ and $c_x$ are constants and
	\begin{align}
		p_t=\frac{\alpha^2-1}{3+\alpha^2}~,~~p_x=\frac{2}{3+\alpha^2}~,~~p_{\psi}=\frac{2\sqrt{2}\alpha}{3+\alpha^2}~.\label{KasnerMtric}
	\end{align}
	
	The Kasner exponents obey $p_t+2p_x=1$ and $p_{\psi}^2+p_t^2+2p_x^2=1$.
	
	\section{The internal periodical structure of black hole}\label{sec4}
	As studied in Refs.~\cite{Zhang:2021vwp,Zhao:2022jvs,Zhao:2023ffs,Zhao:2024jhs,Cao:2024irr,Zhao:2025tqq}, when the extra higher-order nonlinear terms are added to the system, both the condensation behavior and the free energy of the system will change accordingly. However, the aforementioned scenario only considers the influence of the nonlinear terms on the condensation value and does not involve the internal structure of the black hole. In this paper, we will focus specifically on the effect of the nonlinear terms on the internal structure of the black hole.
	
\begin{figure}[t]
		\center
		\includegraphics[width=1\columnwidth]{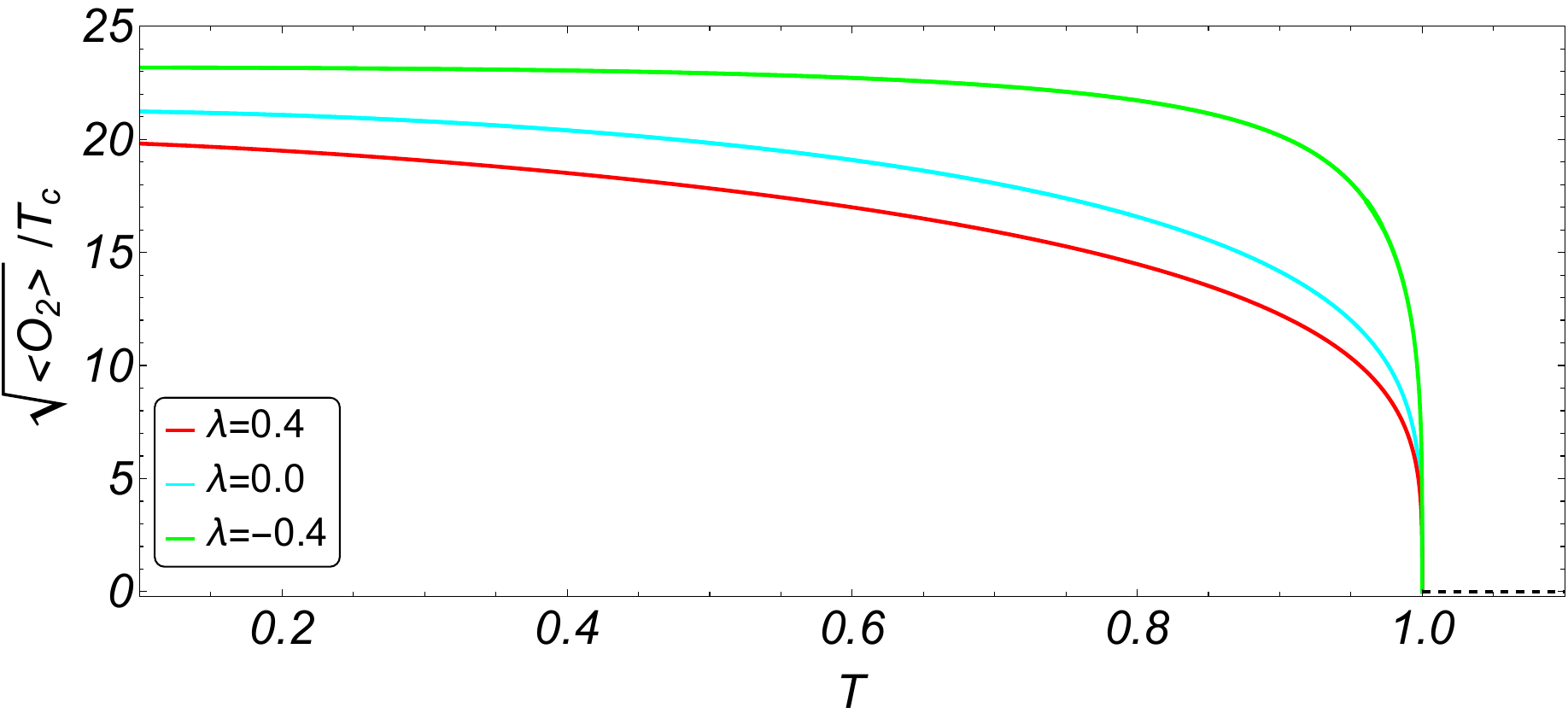}
		\caption{The dependence of the condensates on the nonlinear term $\lambda$ with parameter $\tau=0$. Solid lines represent the condensed solutions, with different colors indicating different values of $\lambda$. The black dashed line represents the normal solution.}\label{CO2andFE2nd}
	\end{figure}

\begin{figure}[!htbp]
		\center
		\includegraphics[width=1\columnwidth]{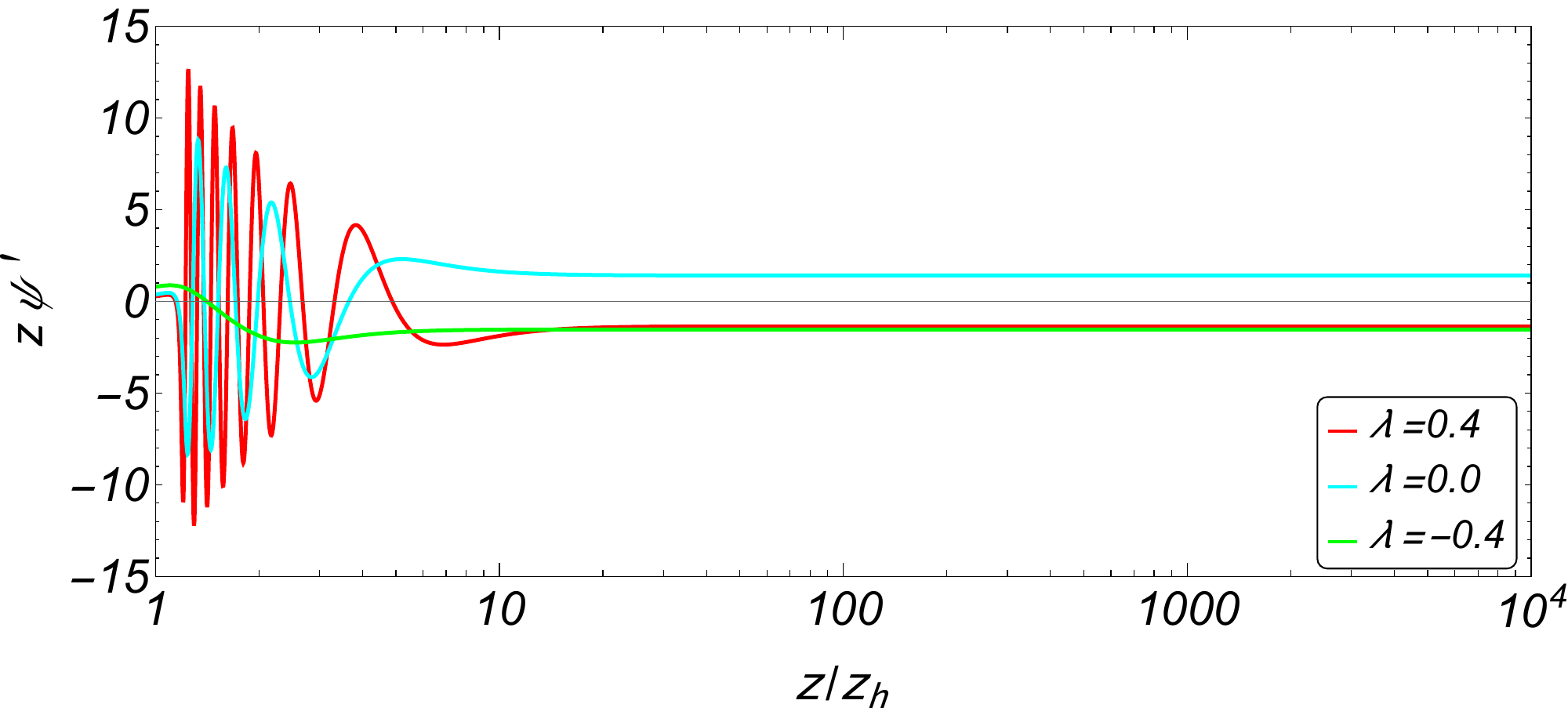}
		\includegraphics[width=1\columnwidth]{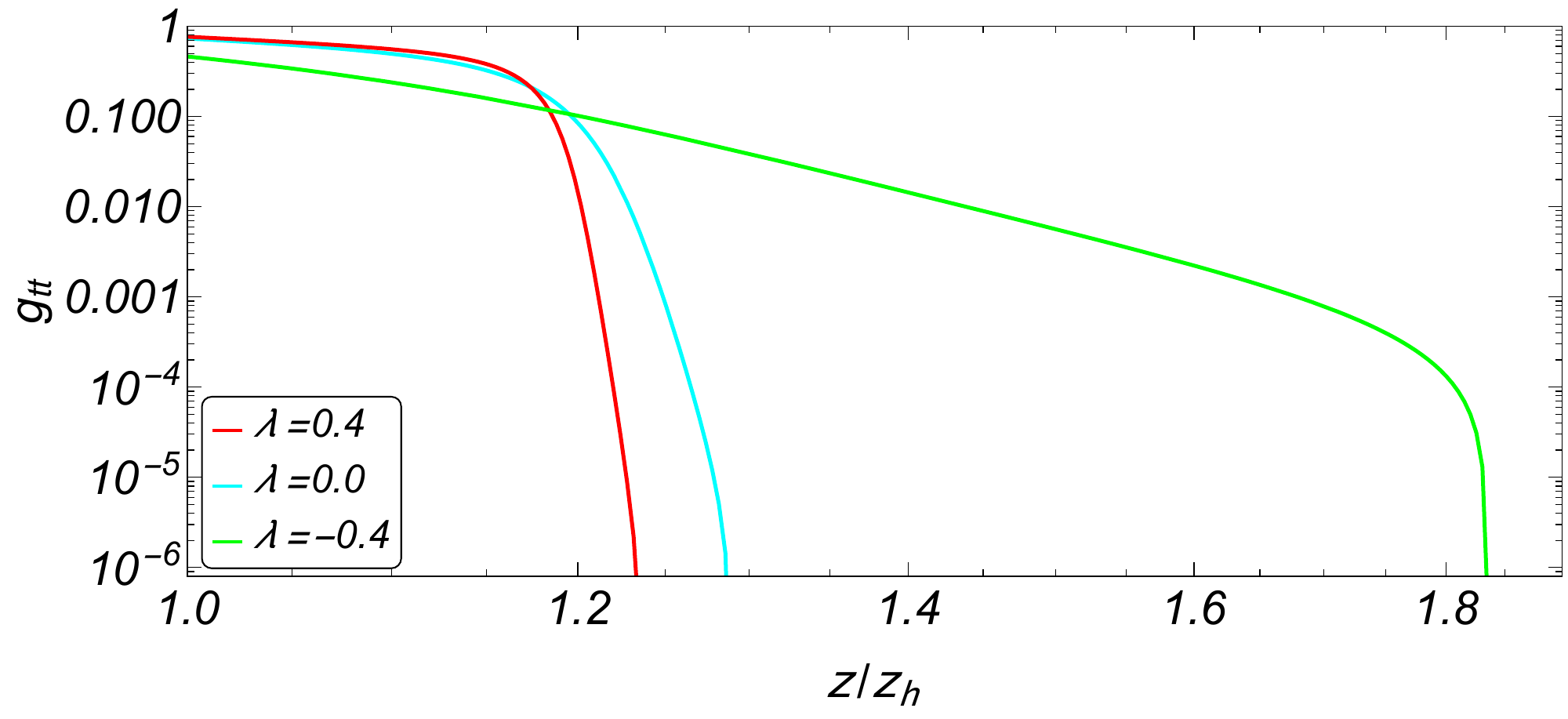}
		\caption{The interior behavior of scalar filed for $T=0.981T/T_c$ with $\lambda=$0.4, 0 and $-0.4$. The upper panel illustrates the complete behavior of the scalar field inside the black hole, while the lower panel corresponds to the collapse of the ER bridge.}\label{interiorZpsi}
	\end{figure}

\begin{figure}[!htbp]
		\center
		\includegraphics[width=1\columnwidth]{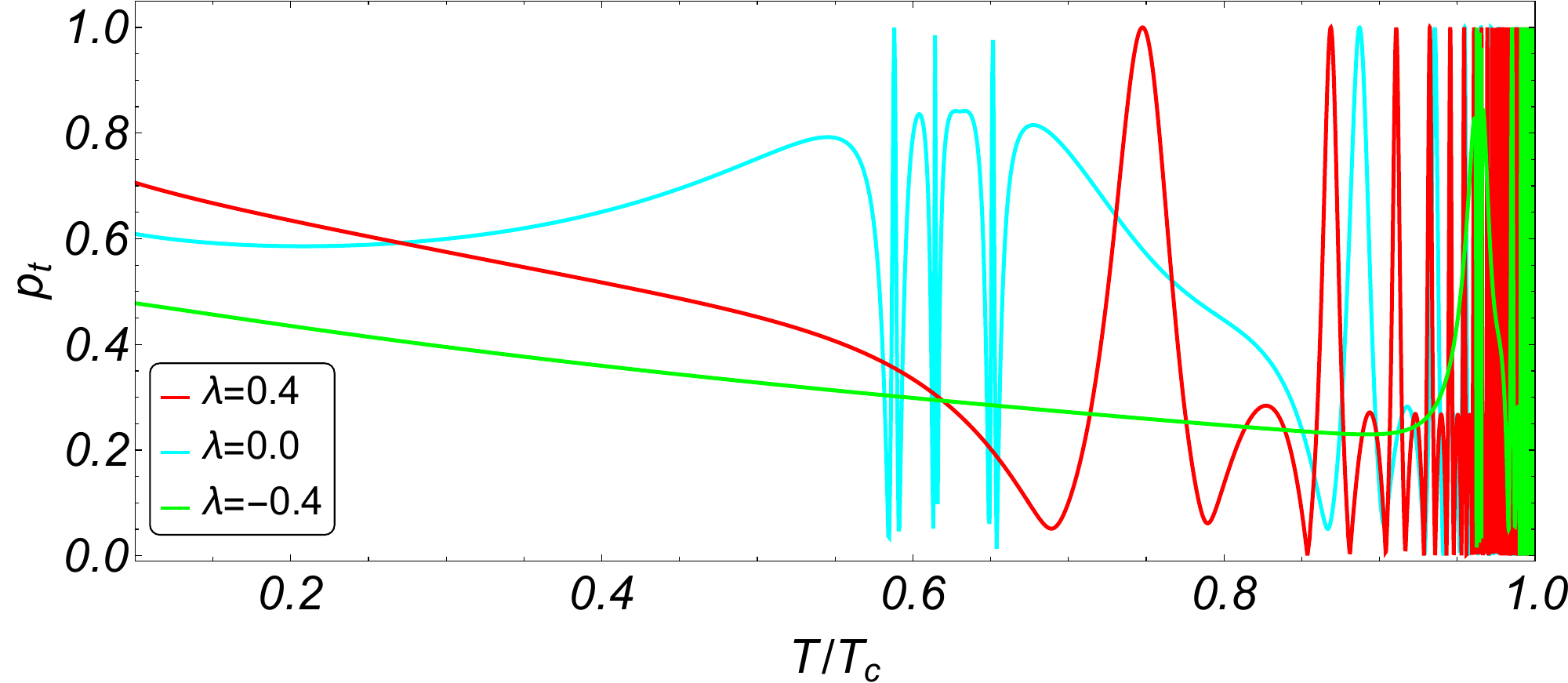}
		\caption{The Kasner exponents $p_t$ as function of temperature $T$. Here, red, blue, and green correspond to $\lambda$ equal to 0.4, 0, and $-0.4$, respectively.}\label{PtTk}
	\end{figure}

	First, we briefly review the influence of the nonlinear term on the system. In Fig.~\ref{CO2andFE2nd}, we present the condensation curves for different values of the parameter $\lambda$ with $\tau=0$. As previously studied in Ref.~\cite{Zhao:2025tqq}, due to the back-reaction effect, even with a negative $\lambda$, the system does not exhibit a zeroth-order phase transition. In Fig.~\ref{CO2andFE2nd}, the influence of the nonlinear term on the condensation curve is limited to modifying the value of the condensate. Meanwhile, in Fig.~\ref{interiorZpsi}, we show the interior solution of the black hole for different values of $\lambda$ in $T=0.981T/T_c$. 
	By computing the Kasner exponents for all temperatures, we obtain the relationship between $p_t$ and temperature $T$, as illustrated in Fig.~\ref{PtTk}.

	\begin{figure}[t]
		\center
		\includegraphics[width=1\columnwidth]{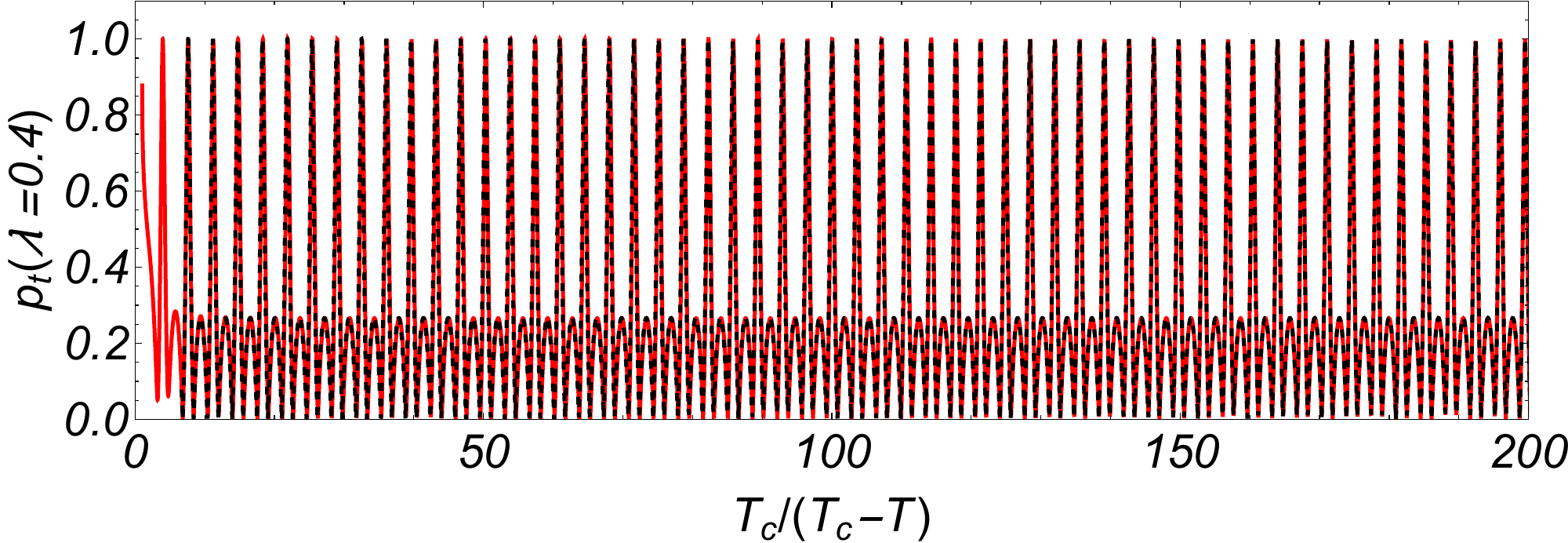}
		\includegraphics[width=1\columnwidth]{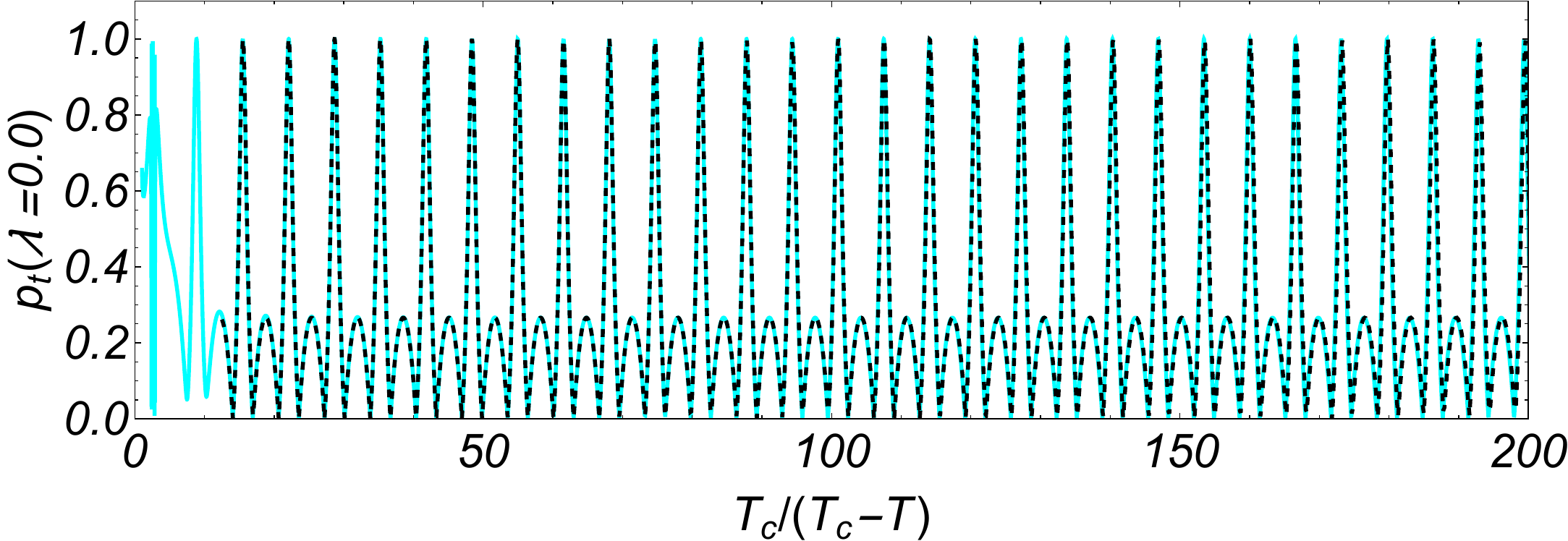}
		\includegraphics[width=1\columnwidth]{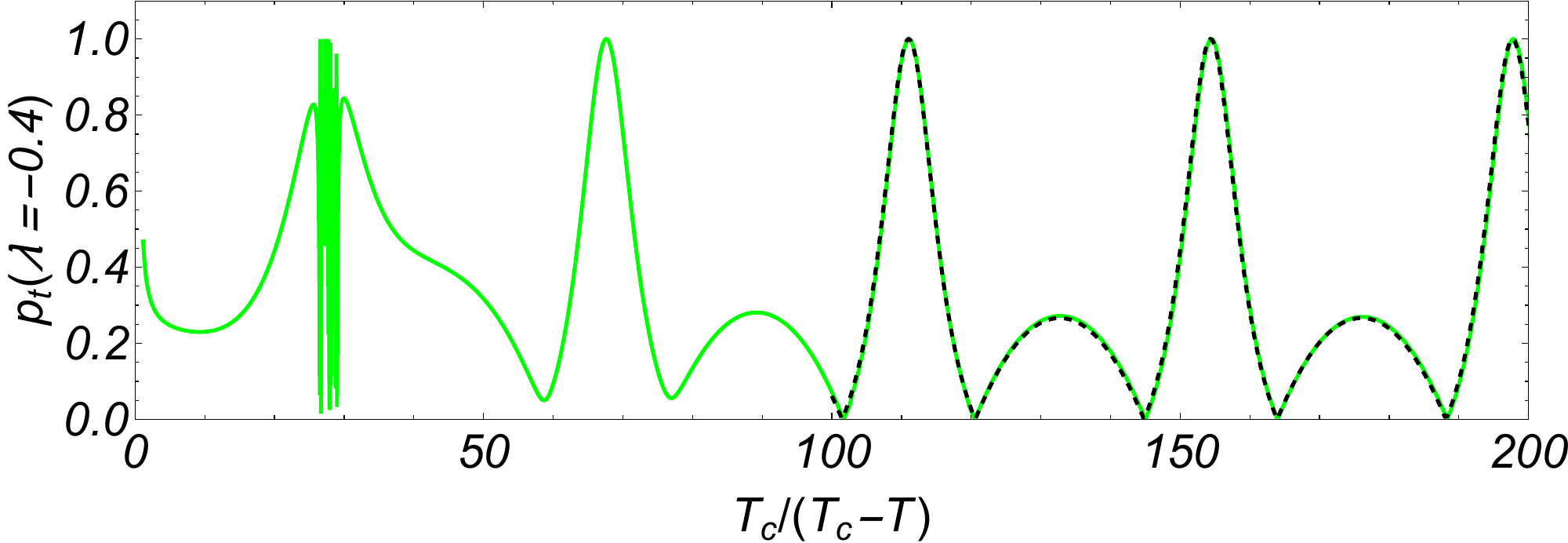}
		\caption{The relationship between the Kasner exponent $p_t$ and $T_c/(T_c-T)$. The solid lines represents the original $p_t$ data, and the black dashed line denote the data derived from $p_t(\alpha_0)$. The parameter values are $\lambda=0.4~(A=1.73923,B=0.884,C=2.7)$, $\lambda=0(A=1.73923,B=0.478,C=2)$, and $\lambda=-0.4~(A=1.85,B=0.0725,C=10.8)$}\label{ptFuliye}
	\end{figure}

	\begin{figure}[!htbp]
		\center
		\includegraphics[width=1\columnwidth]{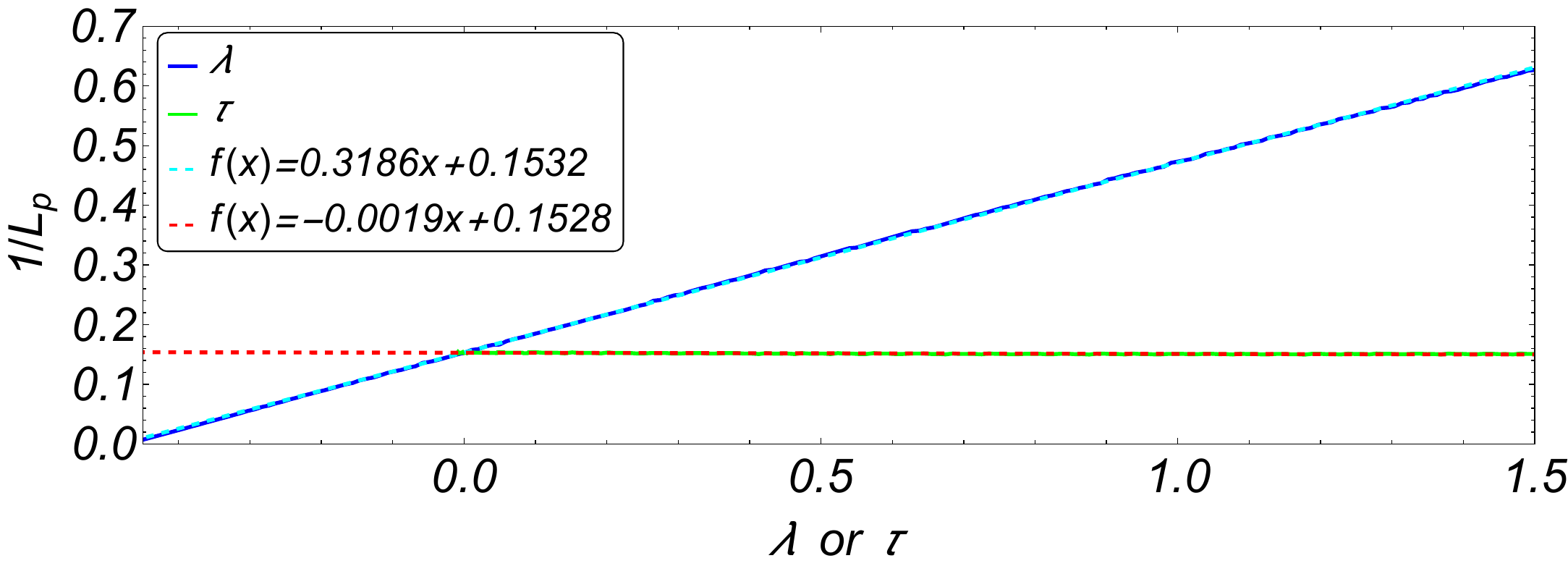}
		\caption{The relationship between periodical length $L_p$ and nonlinear parameters $\lambda$ and $\tau$. The solid lines are the original data, and the dashed lines are the fitting data.}\label{periodicalLambda}
	\end{figure}

	\begin{figure}[!htbp]
		\center
		\includegraphics[width=1\columnwidth]{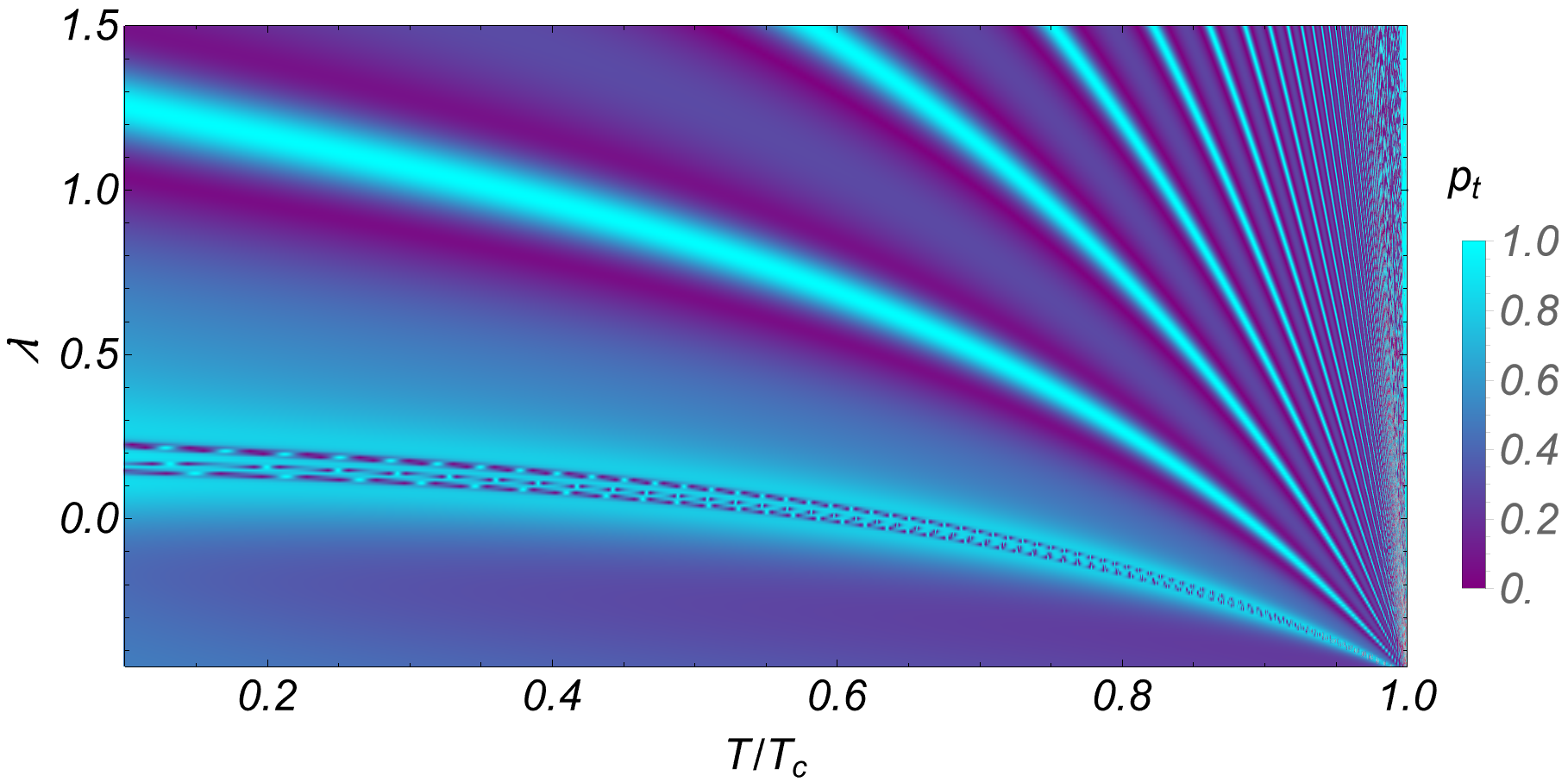}
		\includegraphics[width=1\columnwidth]{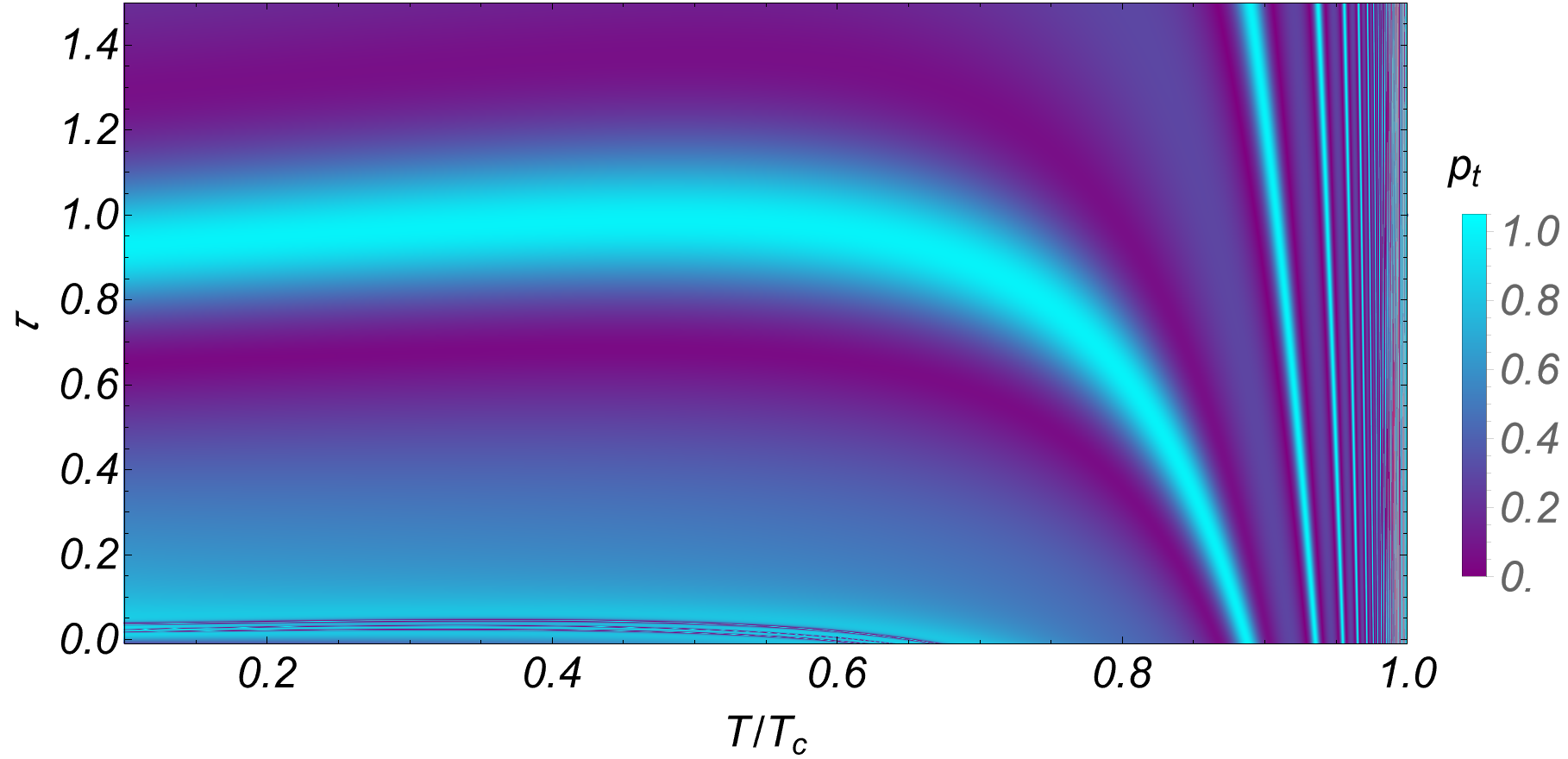}
		\caption{
			Density plots of $p_t$ as a function of the self-interaction parameters and temperature. The color represent the values of $p_t$, as indicated by the color bar on the right.}\label{phaseDiagramlambdaT}
	\end{figure}

	As mentioned in Ref.~\cite{Hartnoll:2020fhc}, the Kasner exponents $p_t$ exhibit highly oscillatory behavior near the critical point. In this paper, by adding a higher-order nonlinear term, we can stretch or compress the interval of this oscillatory behavior. For example, in Fig.~\ref{PtTk}, when $\lambda$=0 (cyan curve), the oscillatory interval is very close to the critical point, and as the temperature decreases, the system will encounter a second oscillatory interval. However, when $\lambda$=0.4 (red curve), the oscillatory interval is significantly enlarged, and the second oscillatory interval will disappear. In contrast, when $\lambda=-0.4$ (green curve), the oscillatory interval is further compressed near the critical point. Therefore, $\lambda$ can be regarded as a magnifying or reducing factor to control the size of the oscillatory region of the Kasner exponents $p_t$.
	
	In addition, this periodic behavior is a function of the form $sin(1/x)$, and its structure can be represented by an analytically solvable expression\cite{Hartnoll:2020fhc}
	\begin{align}
		\alpha_0=-\sqrt{8}\frac{A}{\pi}sin(\frac{B}{1-T/T_c}+C)~.
	\end{align}
	
	
	After performing a transformation on the function $p_t$ in the form of $T/T_c\rightarrow T_c/(T_c-T)$, we obtain a well-defined periodic behavior. It should be noted that the function $\sin(1/x)$ possesses only a single peak. The double-peak oscillatory behavior observed in $p_t$ near the critical point is a consequence of Kasner inversion. This allows us to observe more clearly the influence of the $\lambda$ parameter on the internal structure of the black hole. We present this result in Fig.~\ref{ptFuliye}. Moreover, by extracting the length of each period, we find that the length of the period exhibits a simple linear relationship with $\lambda$. This result is shown in Fig.~\ref{periodicalLambda}, and the relationship can be described by a straightforward linear function $f(x)=ax+b$. 
	
	
	The influence of the parameter $\tau$ on the system differs somewhat from that of $\lambda$. The effect of $\lambda$ is more pronounced in the region near the critical point, while the impact of $\tau$ is more concentrated in the region farther away from the critical point. Similar to the $\lambda$ parameter, we have also calculated the influence of the $\tau$ parameter on the periodic behavior and present the results in Fig.~\ref{periodicalLambda}.

	Finally, we present the relationship between $p_t$ and temperature for different $\lambda$ and $\tau$. In Fig.~\ref{phaseDiagramlambdaT}, we show the density plot of $p_t$ versus $T$ for various $\lambda$ and $\tau$. Distinct stripe structures are clearly observed in the plot. These structures originate from the amplification of the oscillatory region near the critical point by the $\lambda$ and $\tau$. Similarly, the phase diagram clearly shows that increasing the parameter $\lambda$ rapidly expands the oscillatory region near the critical point, while decreasing it compresses this oscillatory region. In contrast, the parameter $\tau$ exhibits only a weak influence on the oscillatory region itself; its effect is more pronounced in lower-temperature region.
	

	\section{Conclusions}\label{sec5}
	This paper systematically investigates the influence of higher-order nonlinear terms on the Kasner geometry inside black holes within a holographic superconductor model. By introducing higher-order nonlinear terms, we numerically compute the behavior of the Kasner exponent $p_t$ as a function of temperature and find that its highly oscillatory structure near the critical temperature can be effectively tuned by the nonlinear coefficients $\lambda$ and $\tau$. Specifically, the parameter $\lambda$ exhibits a notable stretching or compressing effect on the oscillatory region: a positive $\lambda$ amplifies the oscillatory range and makes the periodic structure more pronounced, whereas a negative $\lambda$ compresses it toward the vicinity of the critical point. In contrast, the influence of the parameter $\tau$ is more concentrated in the region away from the critical point.
	
	Through an inverse temperature transformation of the oscillatory interval, this oscillatory behavior manifests as a well-defined periodic pattern. By applying an inverse temperature transformation to the oscillatory interval, this oscillatory behavior manifests as a well-defined periodic pattern. Furthermore, through the extraction of the period length, we find that both parameters $\lambda$ and $\tau$ exhibit clear linear relationships with the period length, offering a novel perspective for understanding the internal structure of black holes. This study extends the understanding of how nonlinear terms affect the dynamical structure inside black holes in holographic superconductors and reveals for the first time their role in tuning the periodic structure of the Kasner exponent. Future research could explore whether similar periodic structures exist in models of different dimensions or other types of holographic superconductors, as well as potential deeper connections between these structures and issues such as black hole information or quantum chaos phenomena.

	\vspace{\baselineskip} 
	\noindent\textbf{Acknowledgements}
	We are grateful to Xin Zhao for useful discussions. This work is supported by National Natural Science Foundation of China (Grant Nos. 12533001, 12575049, 12473001, 12205039, 12305058, 11965013, 12575054 and 12405066). ZYN is partially supported by Yunnan High-level Talent Training Support Plan Young $\&$ Elite Talents Project (Grant No. YNWR-QNBJ-2018-181). This work is also supported by the National SKA Program of China (grant Nos. 2022SKA0110200 and 2022SKA0110203) and the 111 Project (Grant No. B16009).
	
	\vspace{\baselineskip} 
	\noindent\textbf{Data Availability Statement}
	This manuscript has associated data in a data repository. [Authors’ comment: The datasets generated during the current study are available in Zenodo repository, https://doi.org/10.5281/zenodo.18238745]
	
	\vspace{\baselineskip} 
	\noindent\textbf{Code Availability Statement}
	This manuscript has no associated code/software. [Author’s comment: Code/Software sharing not applicable to this article as no code/software was generated or analysed during the current study.]

	\bibliographystyle{apsrev4-1}
	\bibliography{reference}
	
	

	
\end{document}